\documentclass{webofc}
\RequirePackage{lineno}
\usepackage[varg]{txfonts}   
\usepackage{hyperref}
\usepackage{url}
\hypersetup{colorlinks=true,citecolor=black,urlcolor=black,linkcolor=black}
\begin{document}
\title{Imaging shapes of ground-state uranium-238 nuclei in high-energy nuclear collisions at RHIC}
%
%

\author{\firstname{Chunjian} \lastname{Zhang}\inst{1,2}\fnsep\thanks{\email{chunjianzhang@fudan.edu.cn}} \lastname{(for the STAR Collaboration)} 
}

\institute{
Key Laboratory of Nuclear Physics and Ion-beam Application (MOE), and Institute of Modern Physics, Fudan
University, Shanghai 200433, China
\and
Shanghai Research Center for Theoretical Nuclear Physics, NSFC and Fudan University, Shanghai 200438, China
}

\abstract{The shape and orientation of colliding nuclei play a crucial role in determining the initial conditions of the quark-gluon plasma (QGP), which influence key observables such as anisotropic and radial flow. In these proceedings, we present the measurements of $v_2$, $p_{\rm T}$ fluctuations and $v_2-p_{\rm T}$ correlations in $^{238}$U + $^{238}$U and $^{197}$Au + $^{197}$Au collisions at center of mass energies $\sqrt{s_{\rm NN}}=$ 193 and 200 GeV, respectively. Our results reveal significant differences in these observables between the two systems, particularly in the most central collisions. Comparisons with hydrodynamic model calculations indicate a large deformation in the ground states of $^{238}$U nuclei, consistent with previous low-energy experiments. However, data also imply a small deviation from axial symmetry of $^{238}$U~\cite{STAR:2024wgy}. Our work introduces a novel approach for imaging nuclear shapes, enhances the modeling of QGP initial conditions, and sheds light on nuclear structure evolution across different energy scales. The potential applications of this method for other nuclear species are discussed.
}
\maketitle
\section{Introduction}
\label{intro}
In recent decades, the "flow paradigm" of produced particles has been extensively studied in both relativistic heavy-ion experimental measurements at the Relativistic Heavy Ion Collider (RHIC)~\cite{Chen:2024zwk} and the Large Hadron Collider (LHC)~\cite{Song:2017wtw}, as well as theoretical modeling~\cite{Noronha:2024dtq} from large to small systems. These observables have significantly advanced our understanding of initial conditions and transport properties of QGP, a novel state of quantum chromodynamics (QCD) matter. However, fundamental questions and potential pitfalls remain in our understanding of heavy-ion collisions as it requires pushing the boundaries particularly in refining initial conditions of QGP matter~\cite{Jia:2022ozr}. Recent advances in experimental capabilities, integrated with hydrodynamic modeling, provide stringent constraints on initial geometry of QGP and enable us to explore the nuclear shape in yoctosecond-scale ($10^{-24} s$) resolution of nuclear many-body distributions, serving as instantaneous-snapshot $camera$ in the collision dynamics~\cite{Jia_2025}. In particular, collisions of different heavy-ion systems, including $^{238}$U + $^{238}$U, $^{197}$Au + $^{197}$Au, $^{96}$Ru + $^{96}$Ru, $^{96}$Zr + $^{96}$Zr, and $^{16}$O + $^{16}$O, highlight the unique opportunities to decipher nuclear shape in a concerted effort with low-energy approaches, such as density functional theories and the first-principle $ab$ $initio$ frameworks~\cite{Verney:2025efj}.

\section{Flow-assisted nuclear imaging method}
\label{sec-1}
The spatial distribution of nucleons is modeled in a deformed Woods-Saxon (WS) sampling 
\begin{equation}\begin{split}
\label{eq:1}
\rho(r, \theta)=\rho_{0}/\left(1+e^{\left[r-R_0(1+\beta_2({\rm cos}\gamma Y_{2,0}+ {\rm sin} \gamma Y_{2,2}) +\beta_3Y_{3,0}+ \beta_4Y_{4,0})/ a_{0}\right]}\right),
\end{split}
\end{equation}
Here, $a_0$ is the surface diffuseness parameter. and the half-density nucleus radius is given by $R_0=1.2A^{1/3}$. The axial-symmetric quadrupole, octupole and hexadecapole deformation are denoted by $\beta_2$, $\beta_3$, and $\beta_4$, respectively. The deviation from spherical symmetry are described by the spherical harmonics $Y_{l,m}$. The triaxiality parameter $\gamma$ ($0^{\circ} \leq \gamma \leq 60^{\circ}$) determines the relative ordering of the three principal radii of nucleus~\cite{Zhang:2021kxj}.

\noindent Head-on collisions of prolate-deformed nuclei, yielding body-body and tip-tip configurations, enhance the fluctuations in eccentricity and the inverse area of the overlap in the transverse ($xy$) plane, which are effectively transformed into the final-state anisotropies within a linear approximation~\cite{Giacalone:2023hwk,Duguet:2025hwi}. These two extremes lead to enhanced, anti-correlated event-by-event fluctuations in $v_2$ and $[p_{\rm T}]$. We therefore quantify those fluctuations with a simple parametric dependence on shape parameters: $\left\langle v_2^2\right\rangle  =a_1+b_1 \beta_2^2$, $\left\langle\left(\delta p_{\mathrm{T}}\right)^2\right\rangle  =a_2+b_2 \beta_2^2$, and $\left\langle v_2^2 \delta p_{\mathrm{T}}\right\rangle =a_3-b_3 \beta_2^3 \cos (3 \gamma)$, where the positive coefficients $a_n$ and $b_n$ capture the collision geometry
and QGP properties. (For details, see Refs.~\cite{STAR:2024wgy,Giacalone:2021udy,Jia:2021tzt})

\section{Accessing and constraining the $\beta_{\rm 2,U}$ and $\gamma_{\rm U}$ parameters}
\label{sec-2}
\begin{figure*}[ht]
\sidecaption
\centering
\includegraphics[width=0.35\linewidth]{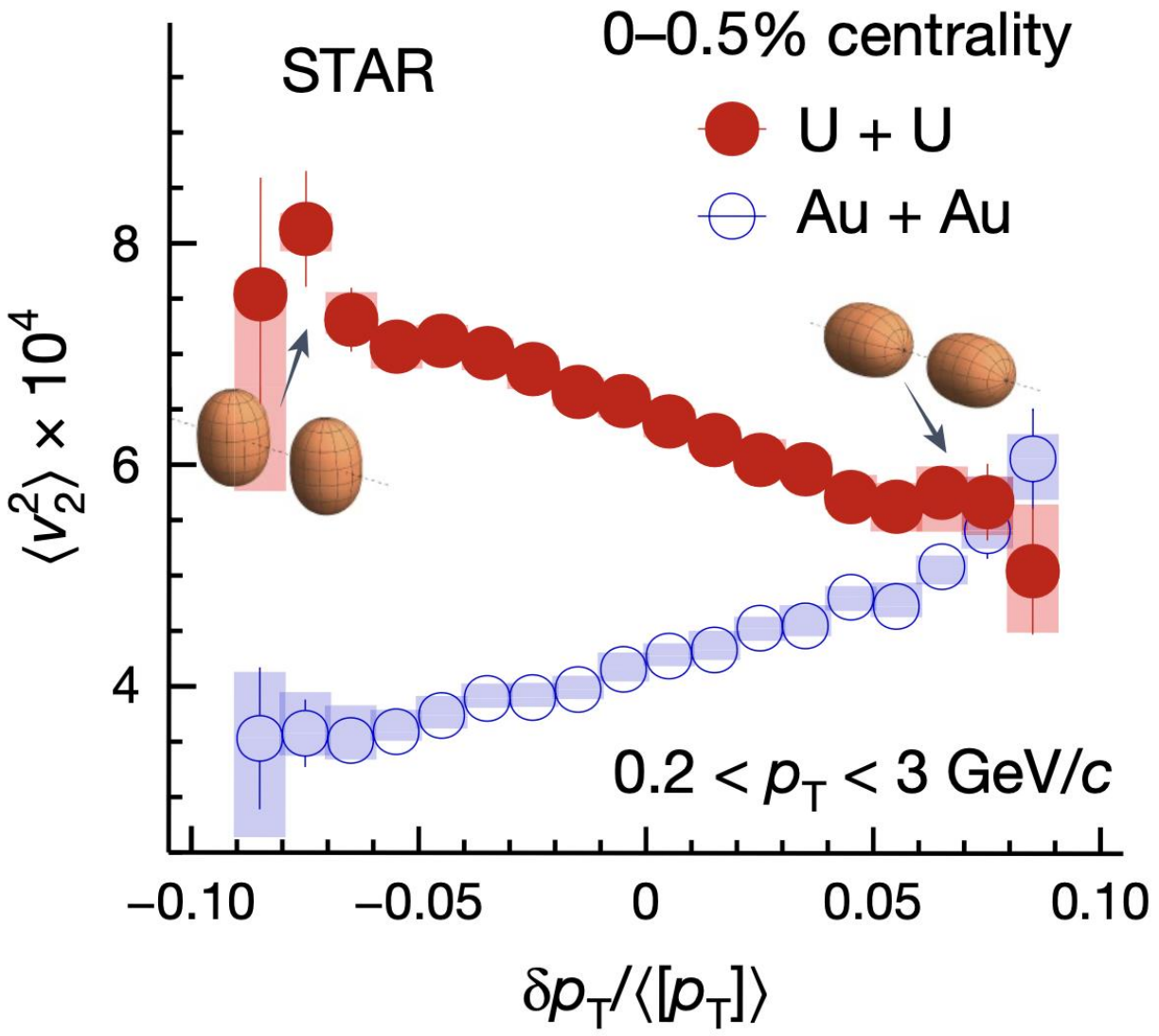}
\caption{$\langle v_2^2 \rangle$ verse $\delta p_{\rm T}/\langle[p_{\rm T}] \rangle$ in 0-0.5\% most central $^{238}$U+$^{238}$U and $^{197}$Au+$^{197}$Au collisions in $0.2 < p_{\rm T} < 3$ GeV/$c$ at $\sqrt{s_{\rm NN}}=$ 193 GeV and 200 GeV, respectively. The head-on body-body and tip-tip configurations are illustrated.}
\label{fig1}     
\end{figure*}

To directly probe the shape–size correlation, we analyze the 0-0.5\% most central collisions and examine the correlation between $\langle v_2^2 \rangle$ and event-wise $\delta p_{\rm T}/\langle[p_{\rm T}] \rangle$ as shown in $^{238}$U+$^{238}$U and $^{197}$Au+$^{197}$Au collisions, as shown in Fig.~\ref{fig1}. A pronounced anticorrelation is observed in $^{238}$U+$^{238}$U collisions confirming the expectations that body–body collisions create a larger, elongated QGP with high $\varepsilon_2$ and small $\delta p_{\rm T}$ values and vice versa for tip–tip
collisions. This enhancement is particularly evident, as $\langle v_2^2 \rangle$ in $^{238}$U+$^{238}$U is approximately twice that in Au+Au at the lowest $\delta p_{\rm T}/\langle[p_{\rm T}] \rangle$ values. However, the two systems exhibit comparable values of $\langle v_2^2 \rangle$ at the higher $\delta p_{\rm T}/\langle[p_{\rm T}] \rangle$. 

\begin{figure*}[t]
\sidecaption
\centering
\includegraphics[width=0.65\linewidth]{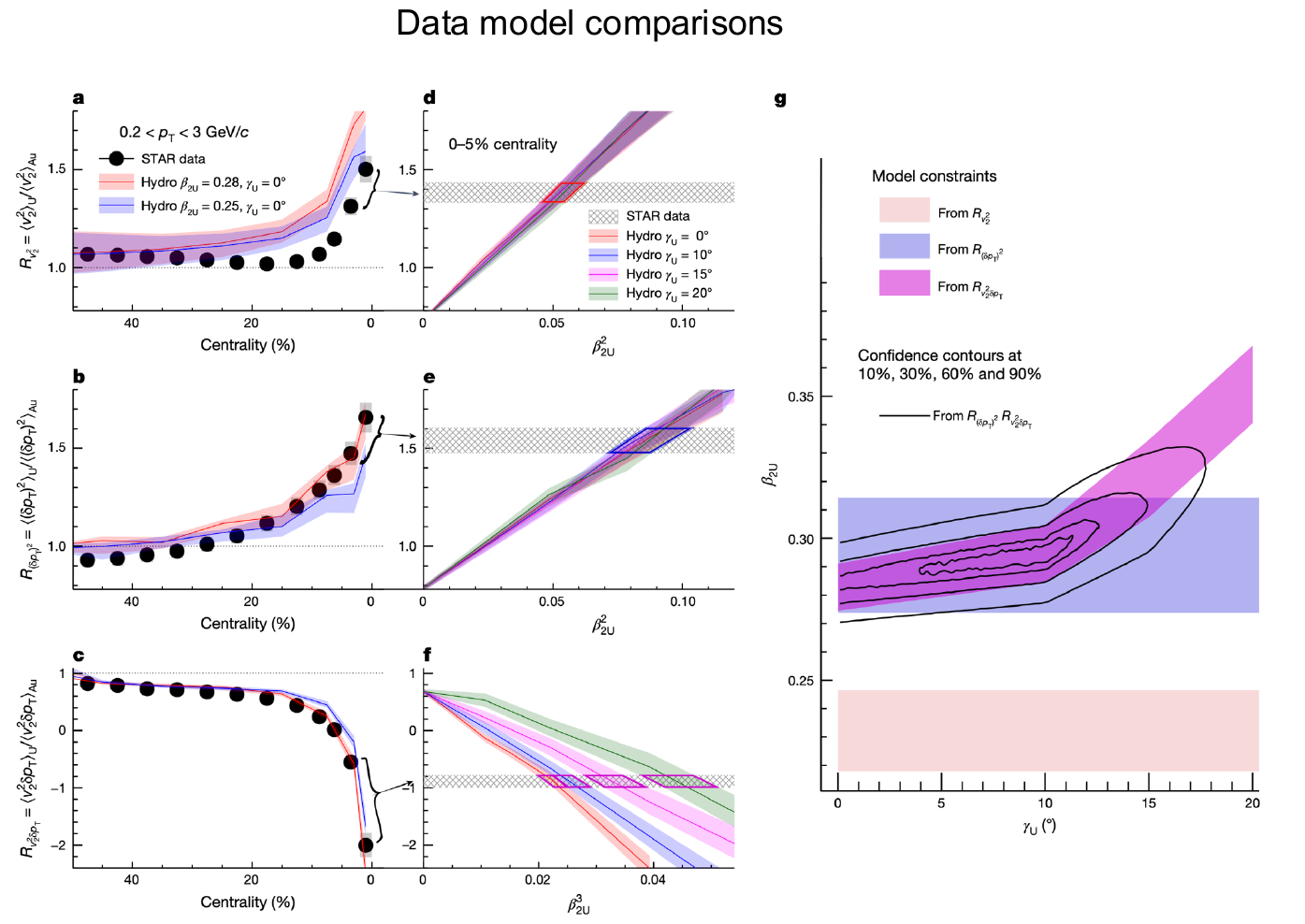}
\caption{Ratios of $\langle v_2^2 \rangle$, $\langle (\delta p_{\rm T})^2\rangle$, and $\langle v_2^2\delta p_{\rm T}\rangle$ in panels a-c between $^{238}$U+$^{238}$U and $^{197}$Au+$^{197}$Au collisions as a function of centrality compared with the hydrodynamic model. The data ratio values in 0-5\% most central collisions are compared with hydrodynamic calculations for different $\beta_{2}$ and $\gamma$ scan in panels d-f. Panel g shows the constraint ranges of ($\beta_{2},\gamma)$ from these three observables separately and the confidence contours obtained by combining $\langle (\delta p_{\rm T})^2\rangle$ and $\langle v_2^2\delta p_{\rm T}\rangle$.}
\label{fig2}      
\end{figure*}

\noindent The final-state effects are largely canceled by taking ratios between two collision systems, leaving model uncertainties mainly from initial conditions~\cite{Zhang:2022fou}. Figure~\ref{fig2}a-c shows ratios of $\left\langle\nu_2^2\right\rangle,\left\langle\left(\delta p_{\mathrm{T}}\right)^2\right\rangle$ and $\left\langle v_2^2 \delta p_{\mathrm{T}}\right\rangle$ between $^{238}$U+$^{238}$U and $^{197}$Au+$^{197}$Au as a function of centrality. $R_{\langle v_2^2 \rangle}$ and $R_{\langle (\delta p_{\rm T})^2\rangle}$ are prominently enhanced in central collisions, demonstrating the geometric role of the large $\beta_{\rm 2,U}$. $R_{\langle v_2^2\delta p_{\rm T}\rangle}$ strongly decrease across centralities, demanding a large $\beta_{\rm 2,U}$ and a small $\gamma_{\rm U}$. Compared to the state-of-the-art IP-Glasma+MUSIC+UrQMD~\cite{Schenke:2020mbo}, calculations with $\beta_{\rm 2,U}=0.28$ provide a good description of $R_{\langle (\delta p_{\rm T})^2\rangle}$ and $R_{\langle v_2^2\delta p_{\rm T}\rangle}$, while $R_{\langle v_2^2 \rangle}$ is slightly overestimated. Figure~\ref{fig2}d-f show that the ratios in the 0-5\% most central range have the greatest sensitivities to the $^{238}$U shape. $R_{\langle v_2^2 \rangle}$ and $R_{\langle (\delta p_{\rm T})^2\rangle}$ exhibit a linear dependence on $\beta_{\rm 2,U}^2$, while the $R_{\langle v_2^2\delta p_{\rm T}\rangle}$ follows a $\beta_{\rm 2,U}^2 {\rm cos}(3\gamma_{\rm U})$ scaling.

\noindent We account for the dominant sources of model uncertainties by varying the QGP transport coefficients and initial conditions, including adjustments in viscosities, nuclear radius $R_0$, skin $a_0$, $\beta_{\rm 2, Au}$, $\gamma_{\rm Au}$ and the higher-order deformations. The intersections between data and model delineate preferred $\beta_{2,\rm U}$ ranges, yielding $\beta_{2 \mathrm{U}}\left(R_{\left(\delta p_{\mathrm{T}}\right)^2}\right)=0.294 \pm 0.021$ and a lower limit value $\beta_{2 \rm U}\left(R_{v_2^2}\right)=0.234 \pm 0.014$. A combined analysis of constraints from $R_{\langle (\delta p_{\rm T})^2\rangle}$ and $R_{\langle v_2^2\delta p_{\rm T}\rangle}$ is performed, yielding $\beta_{\rm 2,U}=0.297 \pm 0.015$ and $\gamma_{\rm U}= 8.5^{\circ} \pm 4.8^{\circ}$ (mean and 1 standard deviation). In addition, comparison with Trajectum model~\cite{Nijs:2023yab} yields constraints of $\beta_{\rm 2,U}=0.275 \pm 0.017$ and $\gamma_{\rm U}= 15.5^{\circ} \pm 7.8^{\circ}$. Combining results from both models gives constraints: $\beta_{\rm 2,U}=0.286 \pm 0.025$ and $\gamma_{\rm U}= 8.7^{\circ} \pm 4.5^{\circ}$. The total systematic uncertainties include an estimate of the non-flow contributions and other sources accounting for detector effects and analysis procedure. More details can be found in Refs.~\cite{STAR:2024wgy,STAR:2025vbp}. The extracted $\beta_{\rm 2,U}$ values are overall consistent with the low-energy studies~\cite{Pritychenko:2013gwa}, while the non-zero $\gamma_{\rm U}$ estimate suggests a slight deviation from axial symmetry.

\section{Summary and applications}
We developed a flow-assisted nuclear shape imaging technique and demonstrated its application as a novel probe of ground-state deformation in heavy $^{238}$U nuclei. This method provides sensitivity for discriminating shape differences between species with similar nuclear mass number ($A$). Our measurements of the simultaneous ratios of $\langle v_2^2 \rangle$, $\langle (\delta p_{\rm T})^2\rangle$, and $\langle v_2^2\delta p_{\rm T}\rangle$ via multi-particle correlations in $^{238}$U+$^{238}$U versus $^{197}$Au+$^{197}$Au collisions reveal a large quadrupole deformation with a small but significant triaxiality in $^{238}$U nuclei. This approach provides crucial constraints for refining the initial conditions of QGP in high-energy nuclear collisions.

\noindent In future, we expect this shape-imaging method would have many possible applications in other isobar or isobar-like species across the nuclear chart from heavy to light nuclei~\cite{Zhang:2021kxj,Ke:2025tyv}. For instance, direct experimental evidences for some exotic nuclear shapes with higher-order deformations remains scarce and challenging to obtain. The recent STAR Collaboration results potentially provide evidence for the presence of modest octupole $\beta_{\rm 3,U}$ and hexadecapole $\beta_{\rm 4,U}$ in $^{238}$U~\cite{STAR:2025vbp,Zhang:2025hvi}. Our technique also provide capabilities to disentangle the complexities of rigid and soft shape fluctuations~\cite{Zhao:2024lpc,Xu:2025cgx,Hagino:2025vxe,Liu:2025fnq}. The fascinating cluster patterns in light nuclei offer a opportunity to bridge the studies of high-energy nuclear collisions and low-energy nuclear structure~\cite{Ma:2022dbh,Huang:2025cjm}. The collision geometry of small systems is particularly sensitive to the nucleon densities and emergent cluster structure~\cite{Summerfield:2021oex,Zhang:2024vkh}. In summary, the collective-flow-assisted nuclear shape-imaging method offers a promising avenue, applicable across collision systems, to unravel the energy evolution of nucleon many-body wavefunctions from low to high energies~\cite{Jia_2025,Giacalone:2025vxa}.

\noindent $\boldsymbol{\rm Acknowledgment:}$ This work is supported in part by the National Key Research and Development Program of China under Contract Nos. 2024YFA1612600, 2022YFA1604900, the National Natural Science Foundation of China (NSFC) under Contract Nos. 12025501, 12147101, 12205051, the Natural Science Foundation of Shanghai under Contract No. 23JC1400200, Shanghai Pujiang Talents Program under Contract No. 24PJA009, the U.S. Department of Energy, Office of Science, Office of Nuclear Physics, under Award No. DE-SC0024602.
\bibliography{reference}{}

\end{document}